\documentclass[prl,twocolumn,groupedaddress,showpacs]{revtex4}
\usepackage{graphicx}
\usepackage{dcolumn}
\usepackage{amssymb,amsmath}

\begin{document}

\title{Competitive Clustering in a Bi-disperse Granular Gas}

\author{Ren\'e Mikkelsen, Devaraj van der
Meer, Ko van der Weele, and Detlef Lohse}
\affiliation{Department of Applied Physics and J.M. Burgers Centre for
  Fluid Dynamics, University of Twente, P.O. Box 217, 7500 AE Enschede, The Netherlands}

\begin{abstract}
A bi-disperse granular gas in a compartmentalized system is
experimentally found to cluster \textit{competitively}: Depending on
the shaking strength, the clustering can be directed either towards
the compartment initially containing mainly small particles, or to the
one containing mainly large particles. The experimental observations
are quantitatively explained within a flux model.
\end{abstract}

\maketitle

Clustering is one of the important features of granular gases,
arising from the inelastic collisions between the particles.
Energy is dissipated in each collision such that a dense region
will dissipate more energy, becoming even denser, resulting in the
formation of a cluster of slow particles
~\cite{Goldhirsch-Jaeger-Kadanoff}. It is a prime example of
structure formation in a system far from equilibrium.

Thus far, most attention has been given to clustering in
mono-disperse systems where all particles are identical. In that
case, the clustering simply occurs in a region that originally is
a bit denser than the others. In the present paper we concentrate
on a bi-disperse granular gas, consisting of large and small
particles. Here the clustering turns out to be competitive: it can
occur either in a region which originally is populated mainly by
{\it small} particles, or in a region with mainly {\it large}
particles.

\begin{figure}
\begin{center}
\includegraphics*[scale=0.6]{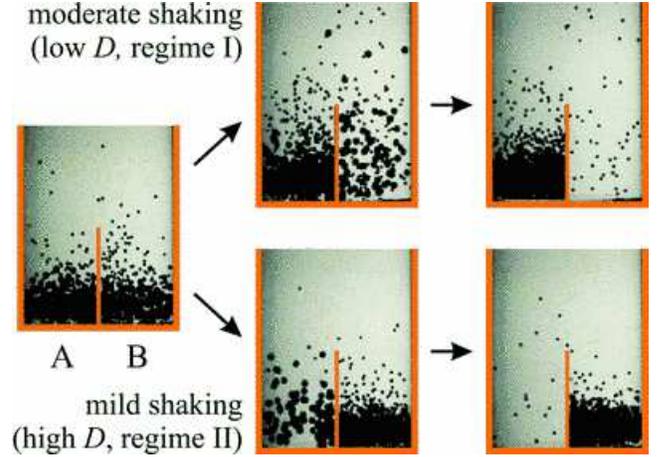}
\caption[]{\small Images from two experiments with a mixture of $P_1$=300 large
  steel beads ($r_1=2.50$ mm, $m_1=0.500$ g) and $P_2$=600 smaller ones
  ($r_2=1.25$ mm, $m_2=0.063$ g). The initial distribution of the beads is the same in
  both experiments: \{$\frac{3}{5}P_1, \frac{1}{3}P_2$\} in
  compartment A and \{$\frac{2}{5}P_1,\frac{2}{3}P_2$\} in
  B. For moderate shaking strength ($f$ = 60.0 Hz, $a$ = 2.0 mm, top
row) a cluster is
  formed in the compartment initially dominated by large
  particles (A). For milder shaking ($f$ = 37.5 Hz, $a$ = 2.0 mm, bottom row) the
  clustering takes place in the compartment initially containing most
  of the small particles (B).}
\label{exp-pic}
\end{center}
\end{figure}

The experimental setup (see Fig.\ref{exp-pic}) consists of a
cylindrical perspex tube with inner diameter 11.2 cm and height
42.2 cm, divided into two equal compartments by a wall of height
6.0 cm. This compartmentalizaton makes it possible to get a
clear-cut picture of the clustering process. The tube is mounted
on a shaker with adjustable amplitude $a$ and frequency $f$. The
inverse shaking strength $D \propto 1/(af)^2$ (defined more
precisely in Eq.(\ref{definition-D}) below) is one of the crucial
parameters for the clustering behavior.

To minimize the effect of statistical
fluctuations in the experiments, we take a sufficiently large number
of beads of each species: $P_1=300$ large and $P_2=600$ small
ones. The small and large beads are both made of steel (so
they have the same coefficient of normal restitution, $e \approx
0.85$), with radius ratio $\psi \equiv r_1/r_2 = 2$.

{\it Clustering behavior}--- The clustering behavior is strongly
influenced by the initial distribution of the two species over the
compartments. As an example we take an initial distribution in
which compartment A has the majority of the large particles, and
compartment B most of the small ones. This situation is depicted
in Fig.\ref{exp-pic}, left, with \{$\frac{3}{5}P_1,
\frac{1}{3}P_2$\} = \{180, 200\} in compartment A and
\{$\frac{2}{5}P_1,\frac{2}{3}P_2$\} = \{120, 400\} in compartment
B. The outcome of the experiment (see Fig.\ref{exp-pic}, right)
depends on the shaking strength.

\begin{figure}
\begin{center}
\includegraphics*[scale=0.3]{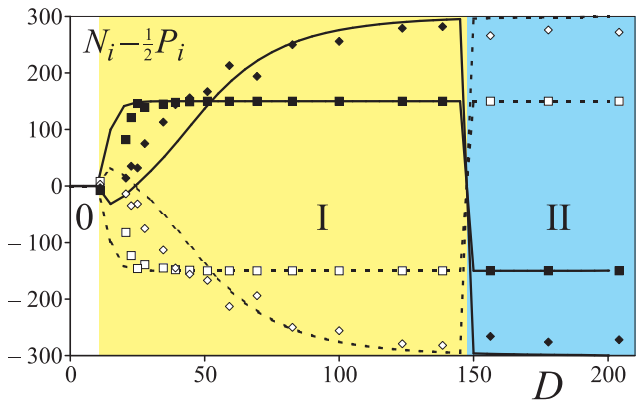}
\caption[]{\small Bifurcation diagram showing the three different
  clustering regimes O, I, and II. The particle numbers $N_i$ ($i = 1,2$) are given relative to the
  symmetric solution: $N_i - \frac{1}{2} P_i$. The squares and
  diamonds are experimental data, corresponding to the numbers of
  large and small particles respectively, in compartment A (solid) and
  B (open). The curves represent the equilibrium situation according to
  the theoretical flux model. The initial condition is always the
  same as in Fig.\ref{exp-pic}.}
\label{bifurc}
\end{center}
\end{figure}

The full bifurcation diagram is given in Fig.\ref{bifurc}: For
sufficiently strong shaking (small $D$) a uniform distribution is
found. This is denoted as regime 0. For moderate shaking strength
(moderate $D$, regime I) the particles cluster in compartment A,
the one initially dominated by large particles. The reason is as
follows: Many of the small beads quickly cluster in compartment A,
where the dissipation is highest due to the greater number of
large beads, which (with their larger mass and surface area) act
as ``coolers''. The remaining beads in compartment B jump higher
than before, since there are fewer collisions, and also large ones
now start to make it over the wall into compartment A. After about
twenty seconds to a minute (depending on the value of $D$) the
final state is reached: a dynamical equilibrium with practically
all large particles and most of the small ones in compartment A,
and only a few rapid small particles in compartment B (see
Fig.\ref{exp-pic}, top row).

For very mild shaking (high $D$, regime II) the clustering process
is much slower and, more importantly, it goes in the
\textit{opposite} direction. The particles now cluster in
compartment B, i.e., in the compartment initially dominated by
small particles (see Fig.\ref{exp-pic}, bottom row). The series of
events is as follows: First, the larger particles hardly jump at
all. They stay close to the bottom, transferring energy from the
vibrating bottom to the smaller particles, which thereby gain
comparatively large velocities (like tennis balls jumping higher
on top of a basketball than on the plain floor). So it is easier
for the small beads to leave compartment A (with more large
beads). The remaining particles become more mobile, and after a
couple of minutes the first large beads start to make it across
the wall into B, where they are immediately swallowed by the
developing cluster. With every particle that leaves compartment A,
the process progressively speeds up. The total clustering time is
in the order of five to twenty minutes, steeply increasing with
the value of $D$.

{\it Flux model} --- The above behavior can be explained in terms of a dynamical flux model. It is a
generalization of the model derived by Eggers for clustering in the
mono-disperse case ~\cite{Eggers,Weele-Meer01,Meer02}. First we consider the
compartments separately, and from this derive the flux of particles
leaving each compartment.

The gas within a compartment is assumed to be in thermal
equilibrium with respect to the granular temperature $T={1\over
2}m_i \left< v_i^2 \right>$ (the mean kinetic energy of the
particles), which is taken to be the same for both species $i =
1,2$. This is an approximation: Two different granular
temperatures may in fact coexist in binary mixtures
~\cite{Losert-Wildman-Feitosa} (and at higher densities the
species are known to segregate ~\cite{Rosato-Knight-Hong}), but
taking this into account would add an unknown and possibly not
constant parameter to our model. As in ref.~\cite{Eggers}, and in
fair agreement with molecular dynamics simulations
~\cite{Mikkelsen-preprint}, we assume that the granular
temperature $T$ is approximately constant throughout the
compartment.

The same simulations ~\cite{Mikkelsen-preprint} show that both
species obey a barometric height distribution (as expected for
dilute granular gases, where one can use the standard equation of
state and momentum balance ~\cite{Eggers}). That is, their number
densities show an exponential decay with the height $z$:
\begin{equation}\label{barometric-distribution}
n_i(z)=n_i(0)\exp(- m_i gz/T).
\end{equation}
The density at ground level follows from $\Omega \int_0^\infty
n_i(z)dz = N_i$ as $n_i(0)= m_ig N_i/\Omega T$. Here $N_i$ is the
number of particles (of species $i$) in the compartment under
consideration and $\Omega$ is its ground area.
Eq.\ref{barometric-distribution} says that the larger particles,
with larger $m_i$, become dilute faster with growing $z$ than the
small particles.

 The temperature $T$ is determined by the balance
between energy input $J_0$ (through collisions with the bottom)
and dissipation $Q$ (through the collisions between particles). To
determine the energy input, we assume for simplicity a
small-amplitude sawtooth motion of the bottom, such that colliding
particles always find it moving upward with velocity $v_b=af$.
This means that a particle coming down with vertical velocity
component $v_{zi}$ is bounced back with $v_{zi}+2v_b$, and the
energy gain per collision is $2m_i v_b(v_{zi}+v_b)$. Multiplying
this by the number of collisions ($\propto n_i(0)v_{zi}$ per
species), and assuming an isotropic Maxwellian velocity
distribution ($\left< v_{zi}^2 \right> = \frac{1}{3} \left< v_i^2
\right>$), this gives the rate of energy input:
\begin{equation}
\label{energy-input-1}
J_0 = \Omega \sum_{i=1,2} n_i(0)\left( \frac{4}{3}v_bT + \frac{2\sqrt 2}{\sqrt
    3} v_b^2 \sqrt{m_iT} \right).
\end{equation}
As typically $v_b \ll v_i$, the first term is much larger than the
second, which we therefore neglect. Then $J_0 = \frac{4}{3}\Omega
(n_1(0)+n_2(0))v_b T = \frac{4}{3} g v_b (m_1 N_1+m_2 N_2)$.

This must be balanced by the rate of energy loss $Q$ through the
inelastic collisions between the particles. It is equal to the
product of the number of collisions per unit volume and the energy
loss per collision ($\propto (1-e^2)[m_i m_j/(m_i +
m_j)][v_i-v_j]^2$), averaged over the velocity distribution, and
integrated over the whole compartment. It consists of three terms,
representing collisions between two particles of type 1, type 2,
and type 1 and 2, respectively ~\cite{note1}:
\begin{eqnarray}
\label{energy-loss}
Q &=& \frac{16 \sqrt\pi g}{\Omega} (1-e^2) T^{3/2}  \times \nonumber \\
&& \hskip-5mm \left( r_1^2 \sqrt{m_1} N_1^2  + r_2^2 \sqrt{m_2} N_2^2 + \frac{4}{7}
  (r_1+r_2)^2 \sqrt{m_{12}} N_1 N_2 \right), \nonumber \\
\end{eqnarray}
\noindent with
\begin{equation}
\label{composite-mass}
\sqrt{m_{12}} = \frac {m_1 m_2
  \left(\frac{m_1}{\sqrt{m_2}}+\frac{m_2}{\sqrt{m_1}} + \frac{3}{4}(\sqrt{m_1}+\sqrt{m_2})\right)}{(m_1+m_2)^2}.
\end{equation}

\begin{figure*}
\begin{center}
\includegraphics*[scale=0.40]{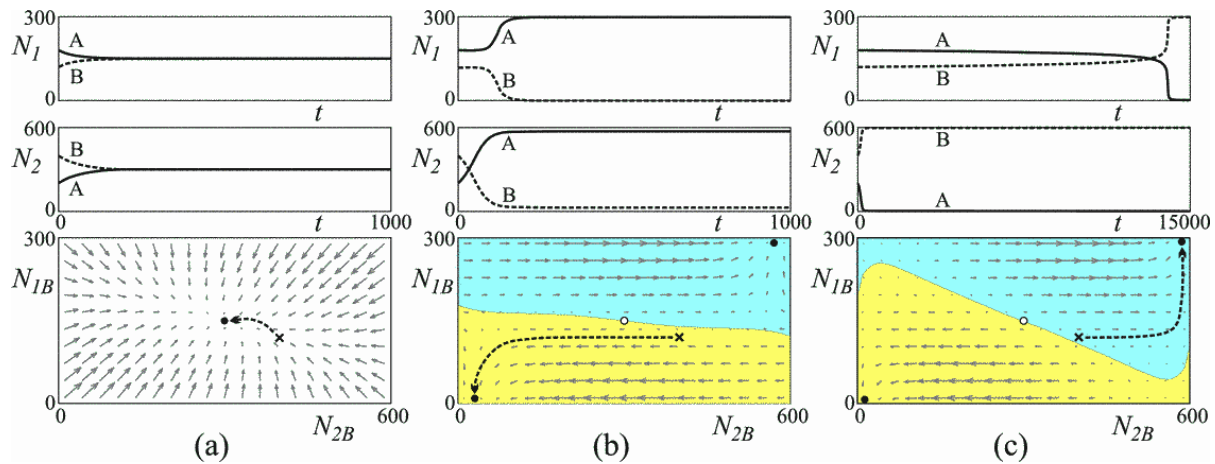}
\caption[]{\small Evolution of the bi-diperse system, calculated
from the
  flux model, for (a) $D=1$, (b) $D=100$, and (c) $D=200$. The top
  rows show the number of large and small particles as a function of
  time (note the different time scales): the solid curves represent
  compartment A, the dashed ones compartment B. The flow diagrams (bottom row)
  show how the contents of B evolve: $N_{1B}$ (large) and $N_{2B}$
  (small). The cross indicates the initial condition \{$N_{1B},
  N_{2B}$\}=\{$120,400$\}. The transition from (a) to (b)
  comes about through a bifurcation, whereas the transition from (b) to (c)
  is seen to be the result of a shifting basin boundary.}
\label{phaseflow}
\end{center}
\end{figure*}

Equating $J_0$ and $Q$ yields the granular temperature $T$
of the compartment:
\begin{equation}\label{temperature}
T = \frac{(af)^2 \mu}{144 \pi (1-e^2)^2},
\end{equation}
\noindent where the effective mass $\mu$ is given by:
\begin{eqnarray}
\label{mu}
\mu(N_1,N_2) &=& \nonumber \\
&& \hskip-1.7cm \left( \frac {\Omega (m_1 N_1 + m_2 N_2)} {r_1^2 \sqrt{m_1} N_1^2 +
    r_2^2 \sqrt{m_2} N_2^2 + \frac{4}{7} (r_1+r_2)^2 \sqrt{m_{12}} N_1
    N_2} \right)^2. \nonumber \\
\end{eqnarray}

\noindent It is through this quantity that the populations of the
two species within one compartment influence each other.

The central quantity of the model is the flux function $F_i$,
i.e., the number of particles (of species $i$) that leave the
compartment per unit time. It is the product of the density
$n_i(z)$ and the horizontal velocity (= $\sqrt{2T/3m_i}$),
integrated over the space above the wall (width $b$) from $z=h$ to
some cut-off height $h+H$ \cite{note2}:
\begin{equation}
\label{fluxfunction}
\begin{aligned}
F_i(N_1,N_2) &= \sqrt{\frac{2T}{3m_i}} b \int_h^H n_i(z)dz \\
&= \sqrt{\frac{2T}{3m_i}} \frac{b N_i}{\Omega} e^{-m_i gh/T}
\left( 1 -
  e^{-m_i g H/T} \right)\\
&= A N_i \sqrt{\frac{m_i}{\mu}} e^{-D m_i/\mu}.
\end{aligned}
\end{equation}
\noindent In the last step we have linearized exp$(-m_i g H/T)$,
implying that $H \ll \left < v_i^2 \right > /g$. The prefactor determining the
absolute rate of the flux is given by
$A=12\sqrt{2\pi/3}(1-e^2)gbH/\Omega af$. The dimensionless parameter
$D$, which governs the clustering behavior, has the form
\begin{equation}
\label{definition-D}
D=144\pi \frac{gh}{(af)^2} (1-e^2)^2.
\end{equation}


The evolution of the number of particles $N_{iA}$ in compartment A
($i=1,2$) is given by the net balance between the (outgoing) flux
from A to B and the (incoming) flux from B to A:
\begin{equation}
\label{flux-ode}
\begin{aligned}
\frac {d N_{iA}}{dt} &= - F_i(N_{1A},N_{2A}) +
F_i(N_{1B},N_{2B}) \\
&= - F_i(N_{1A},N_{2A}) +
F_i(P_1 - N_{1A},P_2 - N_{2A}),
\end{aligned}
\end{equation}
\noindent where we have used particle conservation, $N_{iA}+N_{iB} =
P_i$. The evolution of the (complementary) particle numbers in
compartment B is governed by the same equation with A and B interchanged.

{\it Flux model vs.\ experiment} --- In Fig.\ref{bifurc} the
results from the flux model are shown together with the
experimental data, using the same initial distribution as in
Fig.\ref{exp-pic}. The agreement between theory and experiment is
very good, and the same three regimes are found: For vigorous
shaking (regime 0, $D < 10$) the system settles into the
unclustered homogeneous state. For moderate shaking (regime I,
$10<D<150$) we find clustering in compartment A, and for mild
shaking ($D>150$) the clustering takes place in compartment B. The
transition from regime I to II is quite abrupt.

The experimental observation that the small particles always (in
both regimes I and II) are the first to cluster, followed later by
the large ones, is also reflected in the flux model. This can be
seen in Fig.\ref{phaseflow}, where the evolution of the various
particle numbers (obtained from Eq.\ref{flux-ode}) is shown. The
same plots also illustrate the different timescales for type-I and
type-II clustering: The time it takes the system to reach its
equilibrium situation grows rapidly with growing $D$ (decreasing
shaking strength), just as in experiment.

In the lower part of Fig.\ref{phaseflow} the corresponding flow
diagrams are depicted, showing how the particle numbers $N_{1B}$
and $N_{2B}$ in compartment B evolve, for any initial condition.
The cross denotes the initial condition used in the experiments
(\{$N_{1B}, N_{2B}$\}=\{$120,400$\}), and the arrows show the
evolution of the system. For very strong shaking (left plot) only
one stable fixed point exists: the uniform distribution
\{150,300\}.

In the middle plot, for moderate shaking strength, the
homogeneous state has become unstable and has given way to two new
stable fixed points. These correspond to the compartment B being
either nearly empty (fixed point in the lower left corner) or
well-filled (upper right). Their basins
of attraction are indicated by the shading. Our starting point happens
to lie in the basin of the lower left point, and so we end up with
compartment B being nearly empty. The arrows show that the small
particles cluster first, and only when these have attained their final
distribution do the large ones follow. Indeed, the flux of large particles
is only appreciable when the number of small particles in the
compartment has become very low.

Finally, in the rightmost plot (for mild shaking) we see that the
basin boundary has shifted, which explains the switch from regime
I to regime II: Our initial condition now lies within the basin of
attraction of the fixed point in the upper right corner. So we end
up with a full compartment B. The same plot shows that the fixed
points move further into their corners as $D$ grows, i.e., the
clustering becomes more pronounced for decreasing shaking
strength, just as in the mono-disperse case
~\cite{Eggers,Weele-Meer01}.

\begin{figure}
\begin{center}
\includegraphics*[scale=0.28]{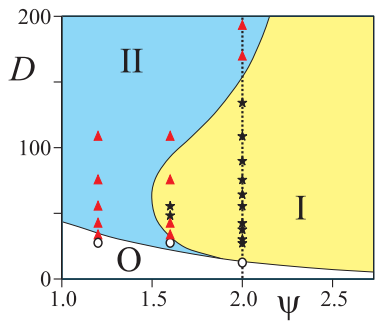}
\caption[]{\small Phase diagram, showing the three clustering
regimes as a function of the inverse shaking strength $D$ and the
size ratio $\psi=r_1/r_2$. The shaded areas are calculated from
the flux model, and the open circles (no clustering), stars (type
I clustering), and triangles (type II clustering) correspond to
experiments. The experimental results on the vertical dashed line
$\psi = 2$ also feature in Fig.\ref{bifurc}. The initial
distribution of
 particles is always taken to be the same as in
 Figs. \ref{exp-pic}-\ref{phaseflow}.}
\label{phasefig}
\end{center}
\end{figure}
{\it Further parameters} --- Until now we have taken the ratio of
the radii $\psi = r_1/r_2 = 2$, but how does the phenomenon depend
on $\psi$ in general? This ratio has a marked effect on the
critical $D$-values where the transitions between the regimes 0,
I, and II take place. In Fig.\ref{phasefig} we show the position
of these regimes as a function of $\psi$, from the flux model
\textit{and} from experiments. The vertical dashed line
corresponds to $\psi = 2$ studied in Figs. 1-3. It is seen that
for $\psi < 1.5$ the transition to regime II is immediate: here
the larger beads are not sufficiently big to compensate for the
fact that they are a minority. It is the larger number of beads
that decide where the cluster goes, just as for the mono-disperse
case ($\psi = 1$). On the other hand, for high values of $\psi$,
the dominant size of the large beads always makes \textit{them}
the decisive factor (only regime I survives). It is precisely the
intermediate region $1.5 < \psi \alt 2.3$ in which the competition
takes place. For $\psi \approx 1.6$ we witness the particularly
interesting sequence 0-II-I-II, both in the model and in
experiment.

Another parameter of interest is the ratio between the total
particle numbers of each species, $\sigma \equiv P_1/P_2$ (up to
now we have always taken $\sigma = 1/2$). Also this ratio
obviously has a big influence on the clustering behavior: a larger
value of $\sigma$ means that the large beads become a more
important minority (or even a majority for $\sigma
> 1$) and hence type-I clustering will gain ground.

{\it Conclusion} --- The clustering behavior of a bi-disperse granular
gas is much richer than the mono-disperse case, and shows a surprising
new feature: The clustering can be directed either towards the
compartment initially containing mainly large particles (type-I
clustering) or to the one containing mainly small particles
(type-II), simply by adjusting the shaking strength. All the experimental
observations can be explained \textit{quantitatively} by a bi-disperse
extension of the Eggers flux model.

{\it Acknowledgments}: This work is part of the research program
of the Stichting FOM, which is financially supported by NWO.
\begin{small}

\end{small}


\begin{thebibliography}{99}

\bibitem{Goldhirsch-Jaeger-Kadanoff} I. Goldhirsch and G. Zanetti, {\it
Phys. Rev. Lett.} {\bf 70}, 1619 (1993); H. Jaeger, S. Nagel, and
R. Behringer, {\it Rev. Mod. Phys.} {\bf 68}, 1259 (1996);
L. Kadanoff, {\it Rev. Mod. Phys.} {\bf 71}, 435 (1999).

\bibitem{Eggers} J. Eggers, {\it Phys. Rev. Lett.} {\bf 83}, 5322
(1999).

\bibitem{Weele-Meer01} K. van der Weele, D. van der Meer, M. Versluis, and
D. Lohse, {\it Europhys. Lett.} {\bf 53}, 328 (2001); D. van der Meer,
K. van der Weele, and D. Lohse, {\it Phys. Rev. E} {\bf 63} 061304 (2001).

\bibitem{Meer02}  D. van der Meer, K. van der Weele, and D. Lohse, {\it Phys. Rev. Lett.} {\bf 88}, 174302 (2002).

\bibitem{Losert-Wildman-Feitosa} W. Losert, D.G.W. Cooper, J. Delour, A. Kudrolli,
and J.P. Gollub, {\it Chaos} {\bf 9}, 682 (1999); R.D. Wildman and
D.J. Parker, {\it Phys. Rev. Lett.} {\bf 88}, 064301 (2002); K.
Feitosa and N. Menon, {\it Phys. Rev. Lett.} {\bf 88}, 198301
(2002).

\bibitem{Rosato-Knight-Hong} A. Rosato, K.J. Strandburg, F. Prinz, and
  R.H. Swendsen, {\it Phys. Rev. Lett.} {\bf 58}, 1038 (1987);
  J.B. Knight, H.M. Jaeger, and S.R. Nagel, {\it Phys. Rev. Lett.}
  {\bf 70}, 3728 (1993); D.C. Hong, P.V. Quinn, and S. Luding, {\it
  Phys. Rev. Lett.} {\bf 86}, 3423 (2001).

  \bibitem{Mikkelsen-preprint} R. Mikkelsen {\it et al.}, preprint
  (University of Twente, 2002).

\bibitem{note1}
We neglect the dissipation resulting from collisions with the
wall, treating them as being completely elastic.

\bibitem{note2}
Above the cut-off height, the state variables of the two
compartments are in equilibrium and hence no net flux occurs. In
principle, $H$ will depend on the mean free path of the particles,
but here we take it to be constant. Naively integrating up to $h
\to \infty$ would lead to an (unphysical) non-zero flux for $N_i =
0$. To see this, consider a compartment without any type 2
particles: then $T \propto 1/N_1^2$ (from Eqs. \ref{temperature}
and \ref{mu}) and hence $F_1$ loses its $N_1$ dependence in the
dilute limit (cf. second line of Eq. \ref{fluxfunction}), making
it non-zero for $N_1 = 0$.

\end{thebibliography}
\end{document}